\documentstyle[12pt]{article}

\author{L.S.F.Olavo\\
Departamento de Fisica, Universidade de Brasilia,\\
70910-900, Brasilia-D.F., Brazil}
\title{Quantum Mechanics as a Classical Theory XV:\\
Thermodynamical Derivation
}

\begin{document}

\maketitle
\begin{abstract}
We present in this continuation paper a new axiomatic derivation of the
Schr\"odinger equation from three basic postulates. This new derivation
sheds some light on the thermodynamic character of the quantum formalism. We
also show the formal connection between this derivation and the one
previously done by other means. Some considerations about metaestability are
also drawn. We return to an example previously developed to show how the
connection between both derivations works.  
\end{abstract}

\vspace{20pt}

\mbox{PACS numbers: 03.65.Bz, 03.65.Ca}

\newpage

\section{Introduction}

In the first paper of this series\cite{eu-1} we showed how to mathematically
derive the Schr\"odinger equation from three basic postulates. At that time
we have made use of an infinitesimal Fourier transformation called by
ourselves the Wigner-Moyal infinitesimal transformation.

The use of this transformation, although very practical for formal
developments
\cite{eu-2}-\cite{eu-14}, implies in some hiding of the underlying physics.

In this continuation paper we will show how to derive the Schr\"o\-din\-ger
equation from more physically sounded axioms. This will not only fix the
derivation itself, but also improve our understanding of the physics
involved, which is our primary goal.

This will be done in the second section. We will show that from three clear
physical postulates it is possible to derive the Schr\"odinger equation.

Then, in section three, we will present the formal (and physical)
connections between the present derivation and the one made in the first
paper of this series\cite{eu-1}.

Section four will be devoted to a brief consideration about the dynamics of
the {\it ensemble statistics}. We will then show that there is a
corresponding Hamilton equation for the {\it ensemble} dynamics.

In section five, we show how our present formulation of the quantum
mechanical postulates enables us to treat metastablility behavior on
qualitative grounds, by means of the entropy concept.

The sixth section will then use the Infinitesimal Transformation formalism
already developed\cite{eu-1} to deal with the metastability problem in a
totally formal fashion.

The last section is devoted to our conclusions.

In the appendix, the basic concepts about the present Schr\"odinger equation
derivation, together with its connection with the previous derivation, will
be exemplified with the concrete problem of a Boltzmann distribution. We
will then show that this probability density has the related probability
amplitudes (wave functions) that are solutions of a Schr\"ondiger equation.
A special application of this last result is then shown for the harmonic
oscillator. This example has already been developed in a previous paper
\cite{eu-12} but now will be used to show that it is a particular
case of a much wider physical framework.

\section{Derivation}

We begin with the Liouville equation 
\begin{equation}
\label{1} \frac{\partial F(x,t;t)}{\partial t}+ \frac{p}{m}\frac{\partial
F(x,p;t)}{\partial x}- \frac{\partial V(x)}{\partial x}\frac{\partial
F(x,p;t)} {\partial p}=0. 
\end{equation}

Integrating this equation with respect to $p$ and using the identities 
\begin{equation}
\label{2} \int F(x,p;t)dp=\rho(x;t)\mbox{ ; } \int
pF(x,p;t)dp=p(x;t)\rho(x;t), 
\end{equation}
where $\rho(x;t)$ is the probability density on the configuration space and $%
p(x;t)$ is the so called macroscopic momentum \cite{Liboff153}, we finally
get 
\begin{equation}
\label{3} \frac{\partial \rho(x;t)}{\partial t}+\frac{\partial}{\partial x}
\left[p(x;t)\rho(x;t)\right]=0, 
\end{equation}
representing a continuity equation for the probability density $\rho(x;t)$.

Now, we may multiply the Liouville equation (\ref{1}) by $p$ and integrate
with respect to this variable to get, using 
\begin{equation}
\label{4} \int p^2F(x,p;t)dp=M_2(x;t), 
\end{equation}
the equation 
\begin{equation}
\label{5} \frac{\partial}{\partial t}\left[\rho(x;t)p(x;t)\right]+ \frac{1}{m%
}\frac{\partial M_2(x;t)}{\partial x}+ \frac{\partial V(x)}{\partial x}%
\rho(x;t)=0. 
\end{equation}

Using the equation (\ref{3}) into equation (\ref{5}) we get, after some
straightforward calculations, the equation 
$$
\frac{1}{m}\frac{\partial}{\partial x}\left[
M_2(x;t)-p^2(x;t)\rho(x;t)\right]+ 
$$
\begin{equation}
\label{6} +\rho(x;t)\left[\frac{\partial p(x;t)}{\partial t}+ \frac{\partial 
}{\partial x}\left(\frac{p^2(x;t)}{2m}\right)+ \frac{\partial V(x)}{\partial
x}\right]=0. 
\end{equation}

The first term to the left may be written as 
$$
M_2(x;t)-p^2(x;t)\rho(x;t)=\int \left[p^2-p^2(x;t)\right]F(x,p;t) dp= 
$$
\begin{equation}
\label{7} =\int \left[p-p(x;t)\right]^2F(x,p;t)dp. 
\end{equation}

We now put 
\begin{equation}
\label{8} \overline{(\delta p)^2} \rho(x;t)=\int
\left[p-p(x;t)\right]^2F(x,p;t)dp 
\end{equation}
and try to find a functional expression for $\overline{(\delta p)^2} $.

To achieve this goal, we begin considering the entropy of this isolated
system as given by 
\begin{equation}
\label{9} S(x;t)=k\ln[\Omega(x;t)], 
\end{equation}
where $\Omega(x;t)$ represents the system accessible states when the
position $x$ varies between $x$ and $x+\delta x$. If this position is free
to vary\cite{Reif290}, then the equal a priori probability postulate says
that 
\begin{equation}
\label{10} \rho(x;t)\propto \Omega(x;t)=e^{S(x;t)/k}. 
\end{equation}

Now, we may ask which probability density $\rho_{eq}(x;t)$, defined upon the
configuration space, is related with the thermodynamic equilibrium of the
system. We are thus interested in the functional variation $\Delta\rho(x;t)$
representing densities variations induced by the fluctuations \cite{Born}.

Then we expand the entropy around a thermodynamic equilibrium configuration,
where the density is given by $\rho_{eq}(x;t)$, as 
\begin{equation}
\label{11} S(x\pm\delta x;t)=S_{eq}(x;t)+\frac{1}{2}\left(\frac{\partial^2
S_{eq}(x;t)} {\partial x^2}\right)_{\delta x=0}(\delta x)^2+..\mbox{  ,} 
\end{equation}
where $S_{eq}(x;t)$ stands for the thermodynamic equilibrium configuration
entropy, given by $\delta x=0$, and where we have already used the fact
that, for this configuration the entropy shall be a maximum and so 
\begin{equation}
\label{12} \left(\frac{\partial S_{eq}(x;t)}{\partial x}\right)_{\delta
x=0}=0. 
\end{equation}

We thus use the fact that 
\begin{equation}
\label{13} \rho_{eq}(x;t)=e^{S_{eq}(x;t)/k}, 
\end{equation}
to get 
\begin{equation}
\label{15} \rho(x,\delta x;t)=\rho_{eq}(x;t) exp\left(-\frac{1}{2k} \left|
\frac{\partial^2 S_{eq}(x;t)}{\partial x^2}\right| (\delta x)^2\right), 
\end{equation}
where we have already used the fact that the second order derivative of the
entropy must be negative near an equilibrium point, together with the
equation (\ref{12}).

Thus, at each point $x$, the probability distribution, with respect to the
small displacements $\delta x(x;t)$, is a gaussian and is related with the
probability of having a fluctuation $\Delta\rho$, on the thermodynamic
equilibrium density, owing to a fluctuation $\delta x$. Equation (\ref{15})
then guarantees that the system will tend to return to its equilibrium
distribution represented by the probability density $\rho_{eq}(x;t)$.

In this case, using the equation (\ref{15}), the mean quadratic
displacements related with the fluctuations are given by 
\begin{equation}
\label{17} \overline{(\delta x)^2}= \frac{\int_{-\infty}^{+\infty}(\delta
x)^2 e^{-\gamma(\delta x)^2} d(\delta x)} {\int_{-\infty}^{+\infty}
e^{-\gamma(\delta x)^2} d(\delta x)}= \frac{1}{2\gamma}, 
\end{equation}
where we have made 
\begin{equation}
\label{17.1} \frac{1}{2\gamma}=k \left|\frac{\partial^2 S_{eq}(x;t)}
{\partial x^2}\right|^{-1}. 
\end{equation}

We have, {\it a priori}, no relation between these displacement fluctuations
and those related with the momenta. We now {\it impose the restriction}
that, in this thermodynamic equilibrium situation, we must have 
\begin{equation}
\label{18} \overline{(\delta p)^2}\mbox{ }\overline{(\delta x)^2}=\frac{%
\hbar^2}{4}. 
\end{equation}
We then get 
\begin{equation}
\label{19} \overline{(\delta p)^2}=-\frac{\hbar^2}{4}\left[ \frac{\partial^2
\ln[\rho(x;t)]}{\partial x^2}\right]. 
\end{equation}

With this last result we may write 
\begin{equation}
\label{20} \overline{(\delta p)^2}\rho(x;t)=-\frac{\hbar^2}{4}\rho(x;t) 
\frac{\partial^2 \ln[\rho(x;t)]}{\partial x^2}. 
\end{equation}

Substituting this expression into equation (\ref{6}) and putting 
\begin{equation}
\label{21} \rho(x;t)=R^2(x;t)\mbox{ ; } p(x;t)=\frac{\partial s(x;t)}{%
\partial x}, 
\end{equation}
we get 
$$
R^2(x;t)\frac{\partial}{\partial x} \left[ \frac{\partial s(x;t)}{\partial t}%
+\right. 
$$
\begin{equation}
\label{22} \left.+\frac{1}{2m}\left(\frac{\partial s(x;t)}{\partial x}
\right)^2+V(x)- \frac{\hbar^2}{2mR(x;t)}\frac{\partial^2 R(x;t)}{\partial x^2%
} \right]=0, 
\end{equation}
which is, as we have already seen\cite{eu-1}, together with the equation (%
\ref{3}), equivalent to the Schr\"odinger equation 
\begin{equation}
\label{23} -\frac{\hbar^2}{2m}\frac{\partial^2\psi(x;t)}{\partial x^2}
+V(x)\psi(x;t)=i\hbar\frac{\partial\psi(x;t)}{\partial t}, 
\end{equation}
with 
\begin{equation}
\label{24} \psi(x;t)=R(x;t)e^{is(x;t)/\hbar}, 
\end{equation}
as we wanted to derive.

One may note that the second expression in the equation (\ref{21}) is
another restriction, since we are imposing that, in our equilibrium
situation, the macroscopic momentum has to be represented by the gradient of
some function.

From what we have said above we may write the new axioms of the theory as:

\begin{itemize}
\item[{\bf A1:}]  Newtonian particle mechanics is valid for every individual
system composing an {\it ensemble};

\item[{\bf A2:}]  The Liouville equation is valid for the description of the 
{\it ensemble} behavior;

\item[{\bf A3:}]  In a thermodynamic equilibrium situation one must have the
restriction (\ref{18}) valid. Besides, the restriction represented by (\ref
{21}) must be also applicable at this equilibrium situation.
\end{itemize}

In the next section we will show how these postulates are connected with the
ones used in the previous derivation\cite{eu-1}.

\section{Connection with Previous De\-ri\-va\-ti\-on}

Within the perspective of our previous derivation\cite{eu-1} we define the
infinitesimal Wigner-Moyal transformation 
\begin{equation}
\label{25} \rho_{eq}(x\pm\delta x/2;t)=Z_Q(x,\delta x/2;t)=\int e^{ip\delta
x/\hbar}F(x,p;t)dp, 
\end{equation}
where we called $Z_Q$ the characteristic function of the probability density 
$F(x,p;t)$.

Clearly 
\begin{equation}
\label{26} \rho(x;t)=\lim_{\delta x\rightarrow 0} Z_Q(x,\delta x/2;t) 
\end{equation}
and 
\begin{equation}
\label{27} p(x;t)\rho(x;t)=\lim_{\delta x\rightarrow 0} -\hbar\frac{\partial
Z_Q(x,\delta x/2;t)}{\partial (\delta x)}. 
\end{equation}

Also 
\begin{equation}
\label{28} \int p^2 F(x,p;t)dp=\lim_{\delta x\rightarrow 0} -\hbar^2\frac{%
\partial^2 Z_Q(x,\delta x/2;t)}{\partial (\delta x)^2}. 
\end{equation}

Then, the right hand side of equation (\ref{7}) may be written as 
$$
\int \left[p-p(x;t)\right]^2F(x,p;t)dp= 
$$
\begin{equation}
\label{29} =\lim_{\delta x\rightarrow 0} \left[ -\hbar^2\frac{\partial^2
Z_Q(x,\delta x/2;t)}{\partial (\delta x)^2}+ \frac{\hbar^2}{\rho(x;t)}\left( 
\frac{\partial Z_Q(x,\delta x/2;t)}{\partial (\delta x)}\right)^2\right], 
\end{equation}
which may be rearranged, after some mathematical calculations, as 
\begin{equation}
\label{30} \overline{(\delta p)^2} \rho(x;t)= -\hbar^2\rho(x;t)\lim_{\delta
x\rightarrow 0} \frac{\partial^2}{\partial (\delta x)^2} \ln\left[Z_Q(x,%
\delta x/2;t)\right]. 
\end{equation}

Performing the expansion of $Z_Q(x,\delta x/2;t)$ in terms of the density,
as given in (\ref{25}), and taking the indicated limit in (\ref{30}), we may
write this last expression, using (\ref{8}) as 
\begin{equation}
\label{31} \overline{(\delta p)^2} \rho(x;t)= -\frac{\hbar^2}{4}\rho(x;t)
\frac{\partial^2 \ln[\rho(x;t)]}{\partial x^2}, 
\end{equation}
which is equivalent to the equation (\ref{20}).

We note that the first two postulates of both derivations are identical \cite
{eu-1}. This means that the formal connection between them has to be settled
upon their third postulates. We thus see that the third postulate we present
in the last section is formally equivalent to postulate the adequacy of the
Infinitesimal Wigner-Moyal Transformation together with the {\it restriction}%
, made in the previous derivation, of writing the characteristic function (%
\ref{25}) as the product 
\begin{equation}
\label{31.a}Z_Q(x,\delta x/2;t)=\psi^{\dag}(x-\delta x/2;t) \psi(x+\delta
x/2;t). 
\end{equation}

In the appendix we show how the connection between both derivations works by
means of a concrete example. It will be seen that the restriction (\ref{31.a}%
), together with the first equality between $\rho_{eq}$ and $Z_Q$, is
equivalent to the impositions made in the third postulate presented in the
last section.

It is also interesting to note that we may consider the characteristic
function $Z_Q(x,\delta x/2;t)$ as a momentum partition function, in the same
mathematical sense that the function $Z$ of usual statistical mechanics is
an energy partition function.

Indeed, we may establish the close formal use of these two functions for the
obtainment of useful statistical quantities. Such a comparison is
represented in Table I, where we present both the cases $Z$ and $Z_Q$ in
contiguous columns.

\section{Statistical Dynamics}

From what we have already said we are now in position to obtain an
interesting equation expressing the dynamical behavior of the system {\it %
statistics}, as considered in terms of the functions $x$ and $p(x;t)$ above
defined\cite{Hermann}.

To achieve this goal we begin considering the mean statistical energy 
\begin{equation}
\label{32} H(x;t)\rho(x;t)=\int\left[\frac{p^2}{2m}+V(x)\right]F(x,p;t)dp. 
\end{equation}
where the mean is taken over the momentum space only. Using the relations (%
\ref{7}), (\ref{8}) and (\ref{19}) we may write 
\begin{equation}
\label{33} H(x;t)=\left[\frac{p(x;t)^2}{2m}+V(x)-\frac{\hbar^2}{8m} \frac{%
\partial^2 \ln\rho(x;t)}{\partial x^2}\right]. 
\end{equation}

In this case, the Liouville equation (\ref{6}) gives 
\begin{equation}
\label{34} \frac{\partial p(x;t)}{\partial t}=-\frac{\partial}{\partial x}
\left[H(x;t)-\frac{\hbar^2}{8m\rho(x;t)} \frac{\partial^2\rho(x;t)}{\partial
x^2}\right], 
\end{equation}
which may be compared with the Hamilton equation related with the individual
systems. We then may call the second term on the right the statistical
potential and our calculations show that it is related with the momentum (or
energy) fluctuations. It is precisely this statistical potential that will
bring the systems to (stable or metastable) thermodynamic equilibrium.

However, we shall once more stress that his equation reflects the dynamics
of the statistical variable $p(x;t)$, since the individual {\it ensemble}
constituents satisfy Newton's equations, as postulated. Indeed, for a
dispersion-free {\it ensemble} we have $p(x;t)=p$. This means that the
dispersion (\ref{8}) is identically zero and we regain Newton equation (in
Hamilton form) from (\ref{34}). We have already shown this property in our
first paper\cite{eu-1} by other ways.

\section{Metastability: Qualitative}

In the second section we found a relation between entropy and the momentum
mean square deviation. This relation was based on the equation (\ref{10})
where the usual connection is made between the entropy and the probability
density.

This relation means, because of the process of derivation, that all the
quantum mechanical solutions we get for a particular problem, the states,
are related with thermodynamic stable or metastable equilibrium
situations---whether we are in an absolute or relative entropy maximum. We
expect that the excited states are related with metastable situations while
the ground state represents stable thermodynamic equilibrium.

To see this qualitatively we first write 
\begin{equation}
\label{m1} S_{n}(x;t)=k\ln [\rho_n(x;t)], 
\end{equation}
where $n$ labels some specific state. It is important to stress that each
particular state will have an associated entropy function.

We may thus define the Gibbs entropy function as 
\begin{equation}
\label{m2}G_n=\int {\rho _n(x;t)\ln [\rho _n(x;t)]dx},
\end{equation}
where here $G_n$ will be more properly a constant, since the probability
densities are time independent---as they are related with equilibrium
situations. This constant, however, will have different values depending on
the specific state under consideration. We thus expect that its values are
such that the Boltzmann theorem applies and the more excited the state is
the most negative the value of $G_n$---for spontaneous transitions between
these states will imply in the decrease of this function values. This last
definition is sufficient to take a glimpse on the process of metastability.

As an application we may consider the harmonic oscillator problem. Thus, we
may use the probability densities related with its states and plot the
corresponding values of the Gibbs entropy function. The result is given in
table II. We note that the metastability condition 
\begin{equation}
\label{m3}\Delta G=G_n-G_m<0, 
\end{equation}
is satisfied when $n>m,n>0$, implying that these states are thermodynamic
metastable ones. If $n=0$ we obviously have a stable thermodynamic
equilibrium. We also see, from table II , that the more excited a state is
the smaller the difference $\Delta G$ with respect to its neighbor $n\pm 1$
states. This may be interpreted as giving the increase in the probability of
transition induced by the fluctuations. It is also interesting to note that
the values of table II have a exponential-decay behavior, as expected.

This qualitative analysis thus suffices to have an account of
meta\-sta\-bi\-li\-ty.

\section{Metastability: Quantitative}

We may deal with the metastability problem directly from the three
fundamental postulates\cite{eu-1}, with the only modification that, when
me\-ta\-sta\-bility is being taken into account, the Liouville equation has
to be replaced by a Master equation. We will use the previous derivation
method throughout\cite{eu-1} because it is the most adequate for formal
calculations.

Thus, we begin by considering that the system is in one of its metastable
equilibrium states, labeled by $i$, and that it is able to make transitions
to some final states $f$. Since we are talking about spontaneous
transitions, we impose that the energy of the initial state be greater than
those of the final ones. The Liouville equation becomes 
\begin{equation}
\label{35} \frac{dF}{dt}=-\sum_{f\ne i}R(i\rightarrow f)F, 
\end{equation}
where 
\begin{equation}
\label{36} \epsilon_i\ge\epsilon_f. 
\end{equation}

Using the same definition of the characteristic function (\ref{25}) and
following the same steps of the demonstration previously done\cite{eu-1}, we
arrive at 
\begin{equation}
\label{37} -i\hbar\frac{\partial Z_Q}{\partial t}-\frac{\hbar^2}{m} \frac{%
\partial^2 Z_Q}{\partial x\partial(\delta x)}+ \delta x\frac{\partial V(x)}{%
\partial x}Z_Q=i\hbar\frac{Z_Q}{\tau}, 
\end{equation}
where we put 
\begin{equation}
\label{38} \frac{1}{\tau}=\sum_{f \ne i}R(i\rightarrow f). 
\end{equation}
Using again\cite{eu-1} the {\it ansatz} 
\begin{equation}
\label{39} Z_Q(x,\delta x/2;t)=\phi^{\dag}(x-\delta x/2;t)\phi(x+\delta
x/2;t), 
\end{equation}
writing $\phi(x;t)$ as in (\ref{24}), expanding (\ref{39}) up to second
order in $\delta x$ and substituting this expansion into equation (\ref{37})
we get, in zeroth and first order in $\delta x$, the two equations 
\begin{equation}
\label{40} \frac{\partial R^2}{\partial t}+\frac{\partial}{\partial x}
\left(R^2\frac{\partial s/\partial x}{m}\right)=-\frac{R^2}{\tau}, 
\end{equation}
and 
\begin{equation}
\label{41} \frac{\partial s(x;t)}{\partial t}+\frac{1}{2m}\left(\frac{%
\partial s(x;t)}{\partial x}\right)^2+V(x)-\frac{\hbar^2}{2mR(x;t)} \frac{%
\partial^2 R(x;t)}{\partial x^2}=0. 
\end{equation}

We note that equation (\ref{40}) represents a continuity equation with a
sink. This equation can be cast into a more interesting appearance if we put 
\begin{equation}
\label{42} R(x;t)=R_1(x;t)e^{-t/\tau}, 
\end{equation}
for, in this case, equation (\ref{40}) may be rewritten as 
\begin{equation}
\label{43} \frac{\partial R_1^2}{\partial t}+\frac{\partial}{\partial x}
\left(R_1^2\frac{\partial s/\partial x}{m}\right)=0, 
\end{equation}
while the equation (\ref{41}) remains the same with the function $R_1(x;t)$
instead of $R(x;t)$ 
\begin{equation}
\label{43.1} \frac{\partial s(x;t)}{\partial t}+\frac{1}{2m}\left(\frac{%
\partial s(x;t)}{\partial x}\right)^2+V(x)-\frac{\hbar^2}{2mR_1(x;t)} \frac{%
\partial^2 R_1(x;t)}{\partial x^2}=0. 
\end{equation}

Equations (\ref{43.1}) and (\ref{43}) are equivalent to the Schr\"ondiger
equation 
\begin{equation}
\label{43.2} -\frac{\hbar^2}{2m}\frac{\partial^2\psi_n(x;t)}{\partial x^2}
+V(x)\psi_n(x;t)=i\hbar\frac{\partial\psi_n(x;t)}{\partial t}, 
\end{equation}
if we put 
\begin{equation}
\label{44} \psi_n(x;t)=R_1(x;t)e^{is(x;t)/\hbar}, 
\end{equation}
and represent the (metastable) thermodynamic equilibrium of the system.

The complete solution to the problem is, thus, 
\begin{equation}
\label{45} \phi_n(x;t)=\psi_n(x;t)e^{-t/\tau}, 
\end{equation}
with the mean lifetime $\tau$ defined as in (\ref{38}). This is the usual
quantum mechanical result found by other means.

\section{Conclusion}

In this paper we have shown how the previous derivation of the
Schr\"o\-din\-ger equation may be accomplished by more sounded physical
principles, which unraveled some of the characteristics of the quantum
formalism---mainly its relation with thermodynamic equilibrium calculations.

The appropriate epistemology underlying all those achievements will be
postponed to a future paper. We hope, however, that this second derivation
process makes the foundational principles of the theory most appealing.

\appendix

\section{Boltzmann Equilibrium}

We want in this appendix to apply some of the concepts developed in the main
text to the special case of a Boltzmann distribution. We will use the
Infinitesimal Transformation approach\cite{eu-1}, since it is more
straightforward, when formalism is being considered.

We thus consider an {\it ensemble} of systems (S) in contact with a
reservoir (O), called the heat bath, responsible to maintaining the
temperature of (S) fixed on some value $T$. The interaction is considered
sufficiently feeble to make it possible to still attach a hamiltonian
function for (S), not depending upon the degrees of freedom of (O). In this
case, in the thermal equilibrium, we have the canonical probability
distribution function given as 
\begin{equation}
\label{a1}F(q,p)=Ce^{-2\beta H(q,p)}, 
\end{equation}
where $H(q,p)$ is the hamiltonian describing the {\it ensemble} and 
\begin{equation}
\label{a2}2\beta =\frac 1{K_BT}, 
\end{equation}
with $K_B$ being the Boltzmann constant, $T$ the absolute temperature and $C$
some normalization constant.

The Hamiltonian may be written as 
\begin{equation}
\label{a3} H(q,p)=\sum_{n=1}^{N}\frac{p^2_n}{2m_n}+V(q_1,..,q_N), 
\end{equation}
where we are assuming that the systems composing the {\it ensemble} have $N$
degrees of freedom, their constituents have masses $m_n,n=1..N$ and are
acted by a potential $V(q_1,..,q_N)$, not depending upon the velocities nor
the time.

Using the Infinitesimal Wigner-Moyal Transformation 
\begin{equation}
\label{a4} Z_Q(q,\delta q/2)=C\int F(q,p)e^{i\sum p_n\delta q_n/\hbar}dp, 
\end{equation}
the characteristic function becomes, after the integration, 
\begin{equation}
\label{a5} Z_Q(q,\delta q/2)=C_1 e^{-2\beta V(q_1,..,q_N)} e^{-\sum\frac{m_n
}{4\beta\hbar^2}(\delta q_n)^2}, 
\end{equation}
where $C_1$ is the constant $C$ modified by the integration process.

This characteristic function is, clearly, a solution to the equation 
\begin{equation}
\label{a6} -\sum_{n=1}^{N}\frac{\hbar^2}{m_n}\frac{\partial^2Z_Q}{\partial
q_n \partial(\delta q_n)}+\sum_{n=1}^{N}\frac{\partial V}{\partial q_n}
(\delta q_n)Z_Q=0, 
\end{equation}
obtained\cite{eu-1} from the Liouville equation by means of the
transformation (\ref{a4}).

As we have already said\cite{eu-1}, it must be possible to write the
characteristic function as the product (see equation (\ref{39})) 
\begin{equation}
\label{a7} Z_Q(q,\delta q/2)=\psi^{\dag}(q-\delta q/2;t) \psi(q+\delta
q/2;t), 
\end{equation}
to derive the Schr\"odinger equation related with the possible thermodynamic
equilibrium distributions. In this case it is easy to see that the
probability amplitudes $\psi$ must be written as 
\begin{equation}
\label{a8} \psi(q;t)=\sqrt{C_1}e^{-\beta V(q_1,..,q_n)}e^{-iEt/\hbar}, 
\end{equation}
giving, for the characteristic function (up to second order as used in the
Infinitesimal Transformation derivation\cite{eu-1}), 
\begin{equation}
\label{a9} Z_Q(q,\delta q/2)=C_1e^{-2\beta\left[V(q_1,..,q_N)+
\frac18\sum(\delta q_n)^2\frac{\partial^2 V}{\partial q_n^2}\right]}. 
\end{equation}

Comparing the expression (\ref{a5}) with (\ref{a8}), we see that the
possibility of writing the characteristic function as the product (\ref{a7})
is related with having, in the vicinity of the point $q=(q_1,..,q_n)$, where
this function is being calculated, 
\begin{equation}
\label{a10} \left.\frac{\partial^2 V}{\partial q_n^2}\right|_{\delta q_n =
0}= \frac{m_n}{\beta^2\hbar^2}. 
\end{equation}

But the expression (\ref{a9}) is equivalent to take 
\begin{equation}
\label{a11} \rho_{eq}(q)=e^{-2\beta V(q_1,..,q_n)}, 
\end{equation}
as the probability density in configuration space and write the
characteristic function as 
\begin{equation}
\label{a12} Z_Q(q,\delta q/2)=\rho_{eq}(q \pm \delta q/2), 
\end{equation}
representing the first equality in (\ref{25}), {\it if, and only if,} 
\begin{equation}
\label{a13} \left. \frac{\partial V}{\partial q_n}\right|_{\delta q_n =0}=0, 
\end{equation}
in the considered point.

The point where conditions (\ref{a10}) and (\ref{a13}) apply defines a
mechanical equilibrium point for the considered systems. That is, the
characteristic function for this problem, where we consider an {\it ensemble}
of systems (S) in thermal equilibrium with a reservoir (O), is equivalent to
the probability density function taken on points slightly apart from the 
{\it mechanical} equilibrium ones. The expression (\ref{a12}) is just the
one we used in section 3, equation (\ref{25}), in the second derivation
method.

Here the connection between the derivations becomes very clear. In our
specific example the entropy is just 
\begin{equation}
\label{a13.a} S(q_1,..,q_N)=-\frac{V(q_1,..,q_N)}{T}, 
\end{equation}
and, since the temperature is fixed, the thermodynamic equilibrium is
attained when (\ref{a13}) and (\ref{a10}) are valid, which means that the
entropy is a maximum---which is also the mechanic equilibrium situation for
this problem. We note that the equation (\ref{a9}) is also equivalent to
equation (\ref{15}).

We may go one step further and ask about the Schr\"odinger equation related
with the solution given by the amplitude (\ref{a8}). In this case,
substituting this amplitude into the Schr\"odinger equation 
\begin{equation}
\label{a14} -\sum_{n=1}^{N}\frac{\hbar^2}{2m_n}\frac{\partial^2\psi}{%
\partial q_n^2} +V(q_1,..,q_N)\psi=E\psi, 
\end{equation}
we get 
\begin{equation}
\label{a15} \sum_{n=1}^{N}\frac{\beta\hbar^2}{2m_n}\frac{\partial^2 V}{%
\partial q_n^2} +V(q_1,..,q_N)-\sum_{n=1}^{N}\frac{\beta^2\hbar^2}{2m_n}
\left[\frac{\partial V}{\partial q_n}\right]^2=E, 
\end{equation}
which, using the relations (\ref{a10}) and (\ref{a13}), becomes, in the
mechanical equilibrium point $q=(q_1^0,..,q_N^0)$, 
\begin{equation}
\label{a16} E=V(q_1^0,..,q_N^0)+NK_B T, 
\end{equation}
as expected for this problem.

The last term on the right is related with the energy given by the reservoir
(O) to the system, for each degree of freedom---and is such as if this
reservoir were constituted of $N$ identical independent harmonic oscillators
contributing, by equipartion of the energy, with the amount $K_B T$. If the
temperature is zero, then it is the same as if the reservoir doesn't exist.
In this case the energy becomes 
\begin{equation}
\label{a17} E=V(q_1^0,..,q_N^0), 
\end{equation}
as should be in a mechanical equilibrium situation.

\subsection{Harmonic Oscillator}

We may particularize our example and consider the case of an {\it ensemble}
of systems comprising $N$ independent harmonic oscillators. We thus have the
potential function of these systems given by 
\begin{equation}
\label{a18} V(q_1,..,q_N)=\frac{1}{2} \sum_{n=1}^{N} m_n \omega_n^2 q_n^2. 
\end{equation}

Equation (\ref{a13}) gives the equilibrium points as $q_i=0,i=1,..N$, while
equation (\ref{a10}) implies that 
\begin{equation}
\label{a19} K_B T=\frac12 \hbar \omega_n, 
\end{equation}
which means that all the frequencies have to be equal.

This last equality gives, for the energy, 
\begin{equation}
\label{a20} E=N\left(\frac{\hbar \omega}{2}\right), 
\end{equation}
and, for the momentum mean square deviation, see equation (\ref{19}), 
\begin{equation}
\label{a21} \frac{\overline{(\delta p_n)^2}}{m_n}=\frac12 \hbar \omega. 
\end{equation}

We may, thus, interpret these results as following: the reservoir (O), when
fixing the systems temperature $T$, gives to each oscillator the amount of
energy represented by (\ref{a19}). These oscillators, even in their mechanic
equilibrium situation will have their momenta and displacements fluctuating,
because of the energy given by the reservoir (O) to attain thermodynamic
equilibrium.

When the temperature approaches absolute zero, the frequencies also
approaches zero---because of (\ref{a19})---signifying that all oscillations
tend to stop. In this case the momentum fluctuations also tend to disappear.
Returning to equation (\ref{15}) we see that there will be a density
fluctuation as small as the temperature is. The probability of having a mean
quadratic displacement obviously also tends to zero. In this case, having
the fluctuations in the momenta and the displacements both tending to zero
violates the thermodynamic equilibrium condition (\ref{18}), and we conclude
that, at this temperature, the oscillators are in mechanic equilibrium but
are not in thermodynamic equilibrium---if they were to be represented by a
Boltzmann distribution function at such a temperature, which is not the
case. Indeed, this last odd result may serve as an indication of the failure
of the {\it ensemble} statistical description by means of the Boltzmann
distribution at such low temperatures.

\newpage

\begin{table}
\begin{center}
\begin{tabular}{|c|c|} \hline
Energy Partition Function &
Momentum ``Partition Function'' \\ \hline \hline
$Z=\sum_r e^{-\beta E_r}$  &
$Z_Q=\int e^{ip\delta x/\hbar}F(x,p;t)dp$ \\ \hline

$\overline{E}=-\frac{\partial ln(Z)}{\partial \beta}$ &
$\overline{p}=\lim_{\delta x\rightarrow 0}-i\hbar\frac{\partial ln(Z_Q)}
{\partial (\delta x)}$ \\ \hline

$\overline{E^2}=\frac{1}{Z}\frac{\partial^2 Z}{\partial \beta^2}$ &
$\overline{p^2}=\lim_{\delta x\rightarrow 0}
-\frac{\hbar^2}{Z_Q}\frac{\partial^2 Z_Q}
{\partial (\delta x)^2}$ \\ \hline

$\overline{(\Delta E)^2}=\frac{\partial^2 ln(Z)}{\partial \beta^2}$ &
$\overline{(\delta p)^2}=\lim_{\delta x\rightarrow 0}
-\hbar^2\frac{\partial^2 ln(Z_Q)}{\partial (\delta x)^2}$ \\ \hline
\end{tabular}
\end{center}
\caption{Comparison between the energy partition function as usually defined 
and the momentum ``partition'' function as defined in the present 
work.}
\end{table}

\begin{table}
\begin{center}
\begin{tabular}{|c|c|} \hline
level(n) &
Gibbs Entropy $(G_n)$ \\ \hline 
$0$  & $-1.07237$ \\
$1$  & $-1.34273$ \\
$2$  & $-1.49859$ \\
$3$  & $-1.60978$ \\
$4$  & $-1.69650$ \\
$5$  & $-1.76803$ \\
$6$  & $-1.82901$ \\
$7$  & $-1.88216$ \\
$8$  & $-1.92927$ \\
$9$  & $-1.97179$ \\
$10$ & $-2.01020$ \\ \hline \hline
\end{tabular}
\end{center}
\caption{Gibbs Entropy values for the harmonic oscillator.}
\end{table}

\end{document}